\documentstyle[epsfig]{mn}

\def\etal{{et al.\ }}
\def\gsim{ \lower .75ex \hbox{$\sim$} \llap{\raise .27ex \hbox{$>$}} }
\def\lsim{ \lower .75ex\hbox{$\sim$} \llap{\raise .27ex \hbox{$<$}} }
\def\msun{\,{\rm M_\odot}}

\def\simlt{\mathrel{\rlap{\lower 3pt\hbox{$\sim$}}\raise 2.0pt\hbox{$<$}}}
\def\simgt{\mathrel{\rlap{\lower 3pt\hbox{$\sim$}} \raise 2.0pt\hbox{$>$}}}
\def\gtrsim{\lower.5ex\hbox{\gsim}}
\def\lesssim{\lower.5ex\hbox{\lsim}}

\def\kms{{\rm\,km\,s^{-1}}}

\def\gcm3{{\rm g\,\, cm^{-3}}}
\def\mbulge{M_{\rm Bulge}}
\def\msunpc3{\msun~{\rm {pc^{-3}}}}
\def\mbh{M_{\rm BH}}
\newcommand{\be}{\begin{equation}}
\newcommand{\ee}{\end{equation}}

\begin{document}

\title[{\it LISA} double black holes]{{\it LISA} double black holes: Dynamics in gaseous nuclear discs}

\author[Dotti, Colpi \& Haardt]{Massimo Dotti$^1$, Monica Colpi$^2$, \& Francesco Haardt$^1$\\
$^1$ Dipartimento di Fisica \& Matematica, 
Universit\'a dell'Insubria, Via
Valleggio  11, 22100 Como, Italy.\\
$^2$ Dipartimento di Fisica "G. Occhialini", Universit\'a di Milano--Bicocca, 
Piazza delle Scienze 3, 20100 Milano, Italy.}

\maketitle \vspace {7cm}

\begin{abstract}

We study the inspiral of double black holes, with masses in the {\it
LISA} window of detectability, orbiting inside a massive
circum--nuclear, rotationally supported gaseous disc. Using
high--resolution SPH simulations, we follow the black hole dynamics in
the early phase when gas--dynamical friction acts on the black holes
individually, and continue our simulation until they form a close
binary.  We find that in the early sinking the black holes lose
memory of their initial orbital eccentricity if they co--rotate with the
gaseous disc.
As a consequence the massive black holes bind forming a
binary with a low eccentricity, consistent with zero within our
numerical resolution limit. 
The cause of circularization resides in
the rotation present in the gaseous background where dynamical
friction operates.  
Circularization may hinder gravitational waves 
from taking over and leading the binary to coalescence.
In the case of counter--rotating orbits the initial 
eccentricity (if present) does not decrease, and the black holes may bind 
forming an eccentric binary.
When dynamical friction has subsided, for equal
mass black holes and regardless their initial eccentricity, angular
momentum loss, driven by the gravitational torque exerted on the
binary by surrounding gas, is nevertheless observable down to 
the smallest scale probed ($\simeq 1$ pc). 
In the case of unequal masses, dynamical
friction remains efficient down to our resolution limit, 
and there is no sign of formation of any ellipsoidal gas
distribution that may further harden the binary.  During inspiral,
gravitational capture of gas by the black holes occurs mainly 
along circular orbits: eccentric orbits
imply high relative velocities and weak gravitational focusing.  Thus,
AGN activity may be excited during the black hole pairing process and
double active nuclei may form when circularization is completed, on
distance--scales of tens of parsecs.

\end{abstract}

\begin{keywords}
Black Hole Physics: binaries, hydrodynamics -- Galaxies: evolution, nuclei -- Gravitational Waves -- Quasar: general
\end{keywords}

\section{Introduction}

Today, we know that a tight link between the formation and evolution
of galaxies and massive black holes (MBHs) exists. Black holes (BHs),
with masses ranging from $\sim 10^6\, \msun$ to $10^9\, \msun$ (e.g.,
Kormendy \& Richstone 1995), are ubiquitous in the centre of nearby
galaxies, and their masses show a tight correlation with the
properties of the host stellar bulges (e.g., Magorrian \etal 1998,
Gebhardt \etal 2000, Ferrarese \& Merritt 2000, H\"aring \& Rix
2004). If MBHs were also common in the past, and if galaxies merge as
implied by hierarchical clustering models of structure formation, then
massive black hole binaries (MBHBs) must have been formed in large
number during the cosmic history.

Close MBHBs are natural, very powerful sources of gravitational
radiation, whose emission is one of the major scientific targets of
the next Laser Interferometer Space Antenna ({\it LISA}; see, e.g.,
Haehnelt 1994, Jaffe \& Backer 2003, Sesana \etal 2005).  The {\it
LISA} interferometer (see Bender \etal 1994), sensitive in the
frequency range between $10^{-5}-10^{-1}$ Hz, will unveil, when in
operation, MBHBs in the mass range $10^3\msun/(1+z)\lsim M_{\rm BH}
\lsim 10^7\msun/(1+z)$: this mass window defines our ``{\it LISA}
black holes".  The lighter objects ($10^{3-5}\msun/(1+z)$) are often
referred to as intermediate mass BHs: their existence has been
conjectured in the framework of hierarchical models of structure
formation (e.g., Volonteri, Haardt \& Madau 2003), and, recently,
gained observational support (van der Marel \etal 2002 for a review,
Gebhardt, Rich \& Ho 2002, Colpi, Possenti \& Gualandris 2002, Gerssen
\etal 2002, Gebhardt, Rich \& Ho 2005, Miller \& Colbert 2004).
Higher mass {\it LISA} BHs are instead in the lower end of the
observed mass distribution, and are the target of our study.

{\it LISA} will detect MBHBs only during the last of a complex
sequence of events that starts when the two MBHs are a few kpc far
apart, and terminates when they reach sub--pc scales, i.e., the
distance at which gravitational waves (GWs) ultimately drive the
final coalescence.  How can MBHs reach the GW emission regime?  The
overall scenario was first outlined by Begelman, Blandford \& Rees
(1980) in their study of the long--term evolution of BH pairs
in dense stellar systems.  They indicated three main processes for the
loss of orbital energy and angular momentum: (a) dynamical friction
against the stellar background acts initially on the BHs as
individual masses, favoring their pairing; (b) MBHs eventually bind to
form a binary when the stellar mass enclosed in the orbit becomes
less than the total mass of the two MBHs. The resulting binary
continues to harden via 3--body interactions with the surrounding stars
until it reaches the separation at which GWs become dominant; (c) in
the third, and last, phase, GW back---reaction shrinks the binary and,
depending on the eccentricity and separation of the orbit, leads to
coalescence.  The dynamical range that a MBHB separation needs to
cover to become a {\it LISA} source is enormous, more than six orders
of magnitude.

Early studies have explored the pairing of MBHs in mergers of purely
collisionless spherical halos (Makino \& Ebisuzaki 1996,
Milosavljevi\'c \& Merritt 2001, Makino \& Funato 2004).  Governato,
Colpi \& Maraschi (1994) first noticed that when two equal mass halos
merge, the MBHs inside their host nuclei are dragged toward the centre
of the remnant galaxy, forming a close pair.  The situation is
different in unequal mass mergers, where the less massive halo is
tidally disrupted, and leaves its MBH wandering in the outskirts of
the main halo.  Thus, depending on the mass ratio and internal
structure of the halos and host galaxies, the transition from phase
(a) to phase (b) can be prematurely aborted.  Similarly, the transit
from phase (b) to phase (c) is not always secured, as the stellar
content of the ``loss cone" may be not enough to drive the binary
separation to the GW emission regime (see, e.g., Milosavljevic \&
Merritt 2001, Yu 2002, Berczik, Merritt \& Spurzem 2005, Sesana,
Haardt \& Madau 2005, in preparation).  In the studies cited above, the background was
purely collisionless.

Since {\it LISA} BH coalescences are likely to be events associated with
mergers of galactic structures at high redshifts, it is likely that
their dynamics occurred in gas dominated backgrounds.  So one might
expect that phases (a) and (b) can be profoundly affected by the
presence of a dissipative component.  Mergers cause large--scale gas
dynamical instabilities that lead to the gathering of cool gas deep in
the potential well of the interacting systems dragging the BHs to the
centre of the remnants.
  
Observations of interacting Luminous Infrared Galaxies (LIRG) in our
local universe have provided evidence of the presence of huge amounts
of cool atomic and molecular hydrogen collected in their cores
(Sanders \& Mirabel 1996).  The total gas mass is
typically $\sim 5 \times 10^9 \msun$, located in the rotationally
supported disc in the inner $\sim 100$ pc (Downes \& Solomon 1998).  
Interestingly, at least in three cases (NGC
6240, Arp 299 and Mkn 463) combined X--ray and infrared observation
hint for the presence of two active MBHs in their nuclei (Hutchings \& Neff 
1988, Komossa \etal 2003, Ballo \etal 2004).

On theoretical ground, the advances in numerical computing allow to
investigate in greater detail the process of BH pairing.  Recently,
Kazantzidis \etal (2005) explored the effect of gaseous dissipation in
mergers between gas--rich disc galaxies with central BHs, using high
resolution N--Body/SPH simulations.  They found that the presence of a
cool gaseous component is essential in order to bring the BH to close
distances, since gas infall deepens the potential well, preserving 
the less massive galaxy against tidal disruption.  Moreover, the
interplay between strong gas inflows and star formation seems to lead
naturally, in these mergers and regardless of the masses of the
interacting galaxies, to the formation, around the two MBHs, of
massive circum--nuclear gaseous discs on a scale $\sim 100$ pc, close to
the numerical resolution limit.  Yet, it is still difficult to
establish the internal kinematic properties of the disc and of the MBH
orbits, whether they are elongated or circular.

The work of Kazantzidis \etal (2005) and the observational evidence
that discs are ubiquitous, have provided our main motivation to study
the process of BH pairing in gaseous circum--nuclear discs.  Escala \etal
(2005, hereinafter ELCM05; see also Escala \etal 2004) 
have studied the role played by gas in affecting the dynamics of
MBHs of equal masses (in a range of $5\times 10^7\msun \leq M_{\rm BH}
\leq 2.5\times 10^9\msun$) moving on circular orbits in Mestel discs
of varying clumpiness.  They explored the decay across phase (a)
controlled by dynamical friction and the transition to the regime (c)
dominated by GW emission, highlighting the role played by
gravitational torques in shrinking the binary.  ELCM05 followed the
orbital decay of a MBHB of total mass equal to $10^8 \msun$ down to
0.1 pc, just around the critical distance for the transition of the GW
domain.                    
The last phase of orbital decay in a gaseous enviroment   
has been studied by Armitage \& Natarajan (2002), and by Milosavljevic \& Phinney (2005). The first 
paper addresses the issue of black hole migration within a pre--exisiting Shakura \& Sunayev (1973) 
accretion disc around a much heavier black hole. It is show how orbital angular
momentum losses due to binary--gas interactions can shrink the orbit in $\sim 10^7$ yrs, down 
to a separation where GW emission rapidly leads the binary to coalescence. The authors suggest 
that an increased accretion rate during the migration is unlikely, while strong, potentially 
observable outflows should preceed the very last phase of the evolution. Milosavljevic \& Phinney (2005) 
argued that, after coalescence, residual disc gas could fill the circum--binary gap, producing, within few 
years, an X--ray {\it LISA} afterglow.     

Further studies are necessary.  First, because {\it LISA} is designed
to detect MBHBs lighter than the mass range explored by ELCM05, and,
second, because, in hierarchical cosmologies, only mergers occurring at
very high redshifts involve almost equal mass MBHBs, while, at later
epochs, coalescences of unequal mass MBHBs are much more common
(Volonteri \etal 2003).  In the present work, we investigate the
sinking process of {\it LISA} BHs inside a massive rotationally
supported disc ($M_{\rm Disc}\sim 10^8\msun$) having finite vertical
extension.  The disc is constructed starting from an initial Mestel
distribution, and is allowed to relax into a new dynamical
equilibrium.  The BHs have equal and unequal masses and are moving
initially on orbits of varying eccentricity: from circular to highly
elongated, in order to reflect different conditions in their sinking
from the kpc--distance scale down to the scale of the nuclear disc
which they inhabit.

We would like to address a number of questions. 

(i) How do eccentric orbits evolve? Do they become circular?  This
issue may be relevant in establishing the initial conditions for the
braking of the binary (phase b) due to the slingshot mechanism and/or
gaseous gravitational torque. Armitage \& Natarajan (2005) have recently shown that 
a residual small (but non zero) eccentricity can be amplified by the binary--disc 
interaction before GW emission dominates. 

(ii) During the sinking process, do the BHs collect substantial
amounts of gas?  This is a query related to the potential activity of
a MBH during a merger and its detectability across the entire
dynamical evolution.

(iii) When a binary forms, how gravitational torques, exerted by the
surrounding gas, depend on the mass ratio between the two BHs?  The
possibility that the binary stalls has to be studied with more
scrutiny and may also depend on the issue (i).

(iv) What mechanisms, besides this torque, can drive the binary into
the GW dominated decaying phase? We consider the effects of BH mass 
growth on their dynamics in \S 5.

The paper is organized as follows. In section \S 2 we describe the
numerical simulations we performed. Results concerning equal mass
binaries are reported in \S 3, while in next \S 4 we describe unequal
mass systems.  Last \S 5 is devoted to discussion and conclusions.

\section{N--Body/SPH simulations}

The circum--nuclear region of a merger remnant is described here as a
superposition of a spherical stellar spheroid and of an axisymmetric
gaseous disc. The disc hosts two BHs treated as collisionless
particles.  The spheroidal component (bulge) is modeled initially as a
Plummer sphere, while the disc as a Mestel distribution.  We evolve
the system using the N--Body/SPH code GADGET (Springel, Yoshida \&
White 2001).
  
\subsection{Bulge and gaseous disc}

The Plummer law for the stellar bulge density is  
\be
\rho (r)={3 \over 4 \pi}{\mbulge\over a^3} \left(1+{r^2\over a^2}\right)^{-5/2},
\ee
where $a$ is the core radius, $r$ the radial
coordinate,  and $\mbulge$ the total mass of the spheroid.

The Mestel disc follows a surface density profile \be \Sigma_{\rm
Disc}(R)={\Sigma_0\,R_0\over R} \ee where $R$ is the radial distance,
projected in the disc plane, and $\Sigma_0$ and $R_0$ reference
values.  The disc is rotationally supported and is characterized by a
circular velocity $V_{\rm cir}$ independent of $R$, in the limit of 
infinitesimal thickness and low temperature. 
The bulge, introduced to stabilize the disc against 
gravitational instabilities, has a mass $M_{\rm Bulge}=6.98 M_{\rm{Disc}}$.

In our current hydrodynamical simulations, the disc has  finite 
radial extension  $R_{\rm Disc}=2a$, and finite vertical thickness
$Z_0$ equal to a tenth of the disc radius $R_{\rm Disc}$.
The vertical density profile is uniform, for a given $R$, initially.
The total disc to  bulge mass ratio inside $R_{\rm Disc}$ is 1:5.
The units of the code are: [Mass]=$10^8\msun$,
[Time]=$2.5\times 10^5$ yr, [Length]=$10.9$ pc, [Velocity]=$41.5 \kms$,
[Density]=$7.8\times 10^4 \msunpc3$. In these units
$V_{\rm cir}=3.7$ (in the limit of zero disc thickness),
and $G=21.9606.$

The equation of state for the gas in the disc is a polytrope,
\be
P=K \rho ^{\gamma}, 
\ee 
where $\gamma$ is equal
to $5/3,$ corresponding to pure adiabatic evolution, and $K$ is set
equal to 2.3325, the value considered by ELCM05, for which the
level of clumpiness during the evolution is minimized.
In internal units $a=5$, $M_{\rm Bulge}=6.98$, $R_{\rm Disc}=10$, 
$M_{\rm Disc}=1$ and the sound speed of gas in the disc at $r=5$ is
$c_s=(\partial P/\partial \rho)^{0.5}=0.289$, corresponding to a temperature
of $T\simeq 10^4 K$. 
In our scheme we have not included the possibility 
that the gas develops a multiphase structure. This has been 
recognized as a relevant feature of self--gravitating discs (see, e.g., 
Wada \& Norman 2001), and it is related to various feed--back mechanisms operating 
in realistic situations (e.g., star formation, radiative heating, shock induced cooling, etc.). 
Though a simple polytropic model, as the one we employed, 
does not catch the detailed thermodynamics of the gas, 
it is, nevertheless, an effective tool to study the orbital decay 
of MBHBs in self--gravitating discs (ELCM05).  

In our simulations, the number of collisionless particles is $10^5$, for the spheroid, while
the number of gas/SPH particles is 235,331. 
With the above figures, our gas mass resolution is $100$ times the mass of a single 
SPH particle, which is, in internal units, $4.25 \times 10^{-6}$ ($425 \msun$ in physical units). 
The mass of bulge stars is $6.98\times 10^{-5}$ ($6980 \msun$), and   
the softening length is 0.1 (1.09 pc), equal for both type of particles. 

\begin{figure}
\vspace{0.5cm}
\centerline{\psfig{file=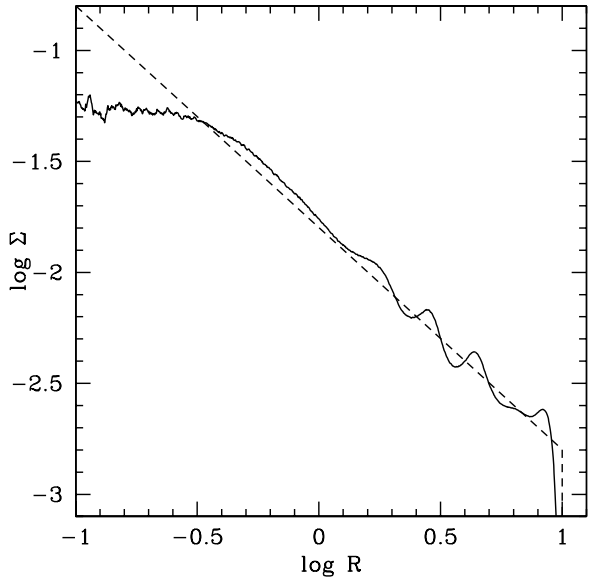,width=3.2in}}
\vspace{-0.0cm}
\caption{\footnotesize Surface density $\Sigma$ as
a function of the distance $R$ from the centre of mass, in internal dimensionless units.
The {\it dashed line} refers to the surface density of the 
Mestel disc, while the {\it solid line} describes  the 
equilibrium profile after an elapsed time $\simeq 10$ (2.6 Myrs),
corresponding to the initial condition of all our simulations.
}
\label{densini}
\vspace{0.5cm}
\end{figure}

Given these initial conditions, the collisionless spherical component,
drawn from the Plummer phase--space distribution function, is in near
equilibrium, while the disc, having finite thickness and homogeneous
vertical density distribution is allowed to evolve into an equilibrium
configuration along the $R$ and $z$--axis.  The disc is evolved for a
dimensionless time of $\sim 10$ (2.6 Myrs) until the density and
pressure fields find their equilibrium.  Initially, the vertical
collapse of the gas increases the pressure gradient in both vertical
and horizontal directions, exciting small waves that propagate
outwards. In settling toward equilibrium, $\Sigma$ modifies and a
small core forms in the central region, as shown in
Figure~{\ref{densini}} where we draw the initial (dashed line) and
equilibrium (solid line) surface density profile; the gas density in
the $z$ direction becomes non uniform, when equilibrium is attained.

\subsection{Black holes}

The BHs are treated as collisionless particles and are placed in the
disc plane. This is an assumption in agreement with the large scale
simulation by Kazantzidis \etal (2005) who find that the BHs pair
inside the massive circum--nuclear disc which forms in the remnant
galaxy, on elongated orbits.  We place our BHs on circular as well as
rather eccentric orbits to bracket uncertainties.

\begin{table}\label{tab:run}
\begin{center}
\caption{Run parameters}
\begin{tabular}{lccccc}  
\hline
\\
run & $M_{\rm BH_1}^*$ & $M_{\rm BH_2}^*$ & $M_{\rm Disc}^*$ & $M_{\rm Bulge}^*$ & $e$\\
\\
\hline
\hline 
\\
A & 1 & 1  &  100  & 698  & 0 \\
B & 1 & 1  &  100  & 698 & 0.97\\
C & 5 & 1  &  100  & 698 & 0 \\ 
D & 5 & 1  &  100  & 698 & 0.95 \\ 
E & 1 & 1  & 0 &  698 & 0.94\\
F$^{**}$ & 5 & 1  &  100  & 698 & 0.95\\
\\
\hline
\end{tabular}\\
\end{center}
\footnotesize{~$^*$ Masses are in units of $10^6 \msun$.}\\
\footnotesize{~$^{**}$ BH$_2$ in run F has a retrograde orbit.}
\end{table}

We consider the case of {\it LISA} BHs
with 1:1 ($10^6\msun-10^6\msun$) 
and 5:1 ($5\times 10^6\msun-10^6\msun$) mass ratio.
The softening of the collisionless BH particles is 
0.1 (1.09 pc). In Table 1  we list the parameters used in our 6
simulations.

\section{Dynamics of equal mass black holes}

\subsection{Circular Orbits}

Run A is aimed to reproduce
run D of ELCM05. Each {\it LISA} BH has a mass $\mbh=0.01M_{\rm
{Disc}}$ and is set on a circular prograde orbit inside the disc. 
The initial separation, relative to the centre of mass, is 
5 (54.5 pc).

\begin{figure}
\vspace{0.5cm}
\centerline{\psfig{file=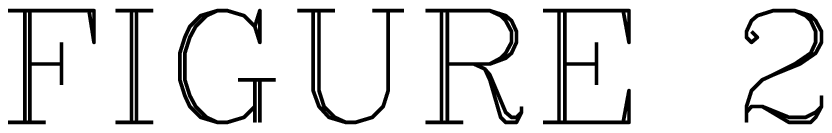,width=3.2in}}

\vspace{-0.0cm}
\caption{\footnotesize Face--on projection of the disc 
for run A at time 2 Myrs. The color coding shows the z--averaged gas density, 
and the bright dots highlight the position
of the two BHs that form prominent wakes behind their
trails.}
\label{circeqden}
\vspace{0.5cm}
\end{figure}

We plot in Figure~\ref{circeqden} the
density map of the gas surrounding the BHs at a selected time,
to show the prominent over-densities that develop behind the holes causing 
their braking. The motion of the BHs is highly supersonic, and this explains the
coherent structure and shape of their wakes (Ostriker 1999). 
Most of the disc gas lying out of BH orbits is somewhat ``squeezed'' into the 
wakes. Qualitatively, the extent of the wakes depends on the amount of disc mass perturbed by 
the orbiting BHs, which is a function of the BH masses.

\noindent

\begin{figure}
\vspace{0.5cm}

\centerline{\psfig{file=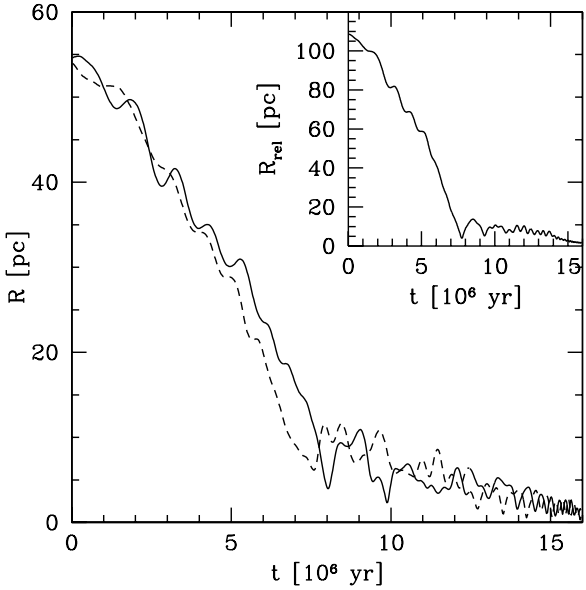,width=3.2in}}

\vspace{-0.0cm}
\caption{\footnotesize Distance $R$ from the
centre of mass, as a function of 
time $t$, for equal mass BHs in run A. {\it Solid} and {\it
dashed} lines refer to the different BHs.  The insert
gives the BH relative separation $R_{\rm rel}$ 
versus time. Integration is halted when the resolution limit ($\simeq 1$ pc)
is attained. }
\label{circeq}
\vspace{0.5cm}
\end{figure}

Figure~\ref{circeq} shows the inspiral of the two BHs sinking
because of dynamical friction mainly due to the gas component. 
The BHs evolve maintaining their orbits nearly circular until
they reach the central, numerically unresolved, region.  The time evolution of the BH 
relative separation $R_{\rm rel}$ is
plotted in the inset of Fig.~\ref{circeq}, and 
is in agreement with ELCM05.

\begin{figure}
\vspace{0.5cm}

\centerline{\psfig{file=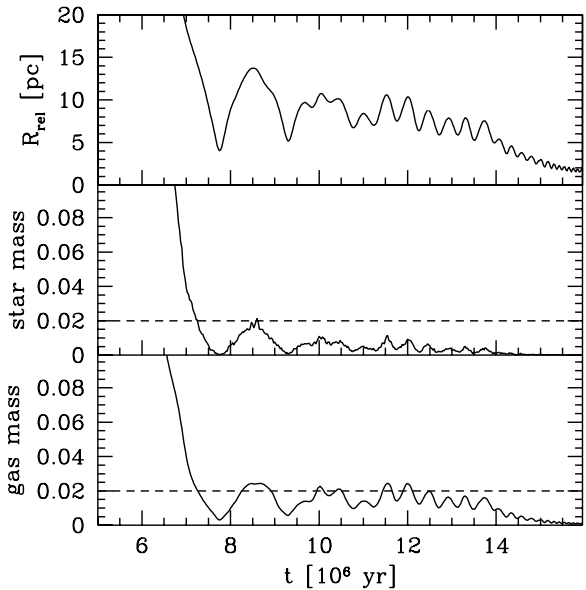,width=3.2in}}

\vspace{-0.0cm}
\caption{\footnotesize 
Upper panel: relative distance $R_{\rm rel}$
between the BHs versus time, for run A. 
Middle (lower) panel: mass (in internal units)
in stars (gas) enclosed inside $R_{\rm rel}$, against time. 
Horizontal dashed lines indicate the total mass of the BHs.}
\label{dfcirceq}
\vspace{0.5cm}
\end{figure}

Before reaching the force resolution limit, where integration stops,
orbital decay becomes less efficient since the nature of the drag
changes.  Figure~\ref{dfcirceq} gives, at late times, the mass in star
and gas enclosed inside the sphere defined by $R_{\rm rel}$.  When
such mass is $\simgt 2\,M_{\rm BH},$ dynamical friction acts on the
BHs as individual objects. At a time $t\simeq 8$ Myrs the mass drops below 
$2\,M_{\rm BH}$ and dynamical friction becomes inefficient.
The BHs now are bound in a ``binary'', and the orbital decay proceeds further,
but at a lower pace. This corresponds to the ``transition regime'' defined
by ELCM05. Our resolution limit does not permit to test the 
subsequent ``ellipsoidal regime'' (ELCM05). 
We note that an ellipsoidal distribution of gas is already present when the binary is
formed, as shown in Figure~\ref{isodens}.

The BHs maintain a nearly circular orbit and this implies that the
relative velocity between the BHs and the rotationally supported gas
particles is small. This suggests that some of the gas in the wake may
become bound to the BHs while spiraling inwards. We find that the
mass in gas particles associated to the over-density
centred around each BH is $\sim 20\%$ of its mass, during inspiral.
This number is calculated by summing the masses of 
all gas particles associated with a spherical density 
excess measured relative to the unperturbed disc, at the 
same position. This provides only an approximate estimate of the mass that
can become bound to the BHs.
The poor resolution close and inside the BH sphere of influence, and the 
simple thermodynamics used, 
prevent us from giving an estimate of the mass accretion rate on
the horizon distance-scale.
We can only speculate that in this phase the BH may accrete, generating
a double nucleus AGN.

\subsection{Eccentric orbits}

\begin{figure}
\vspace{0.5cm}

\centerline{\psfig{file=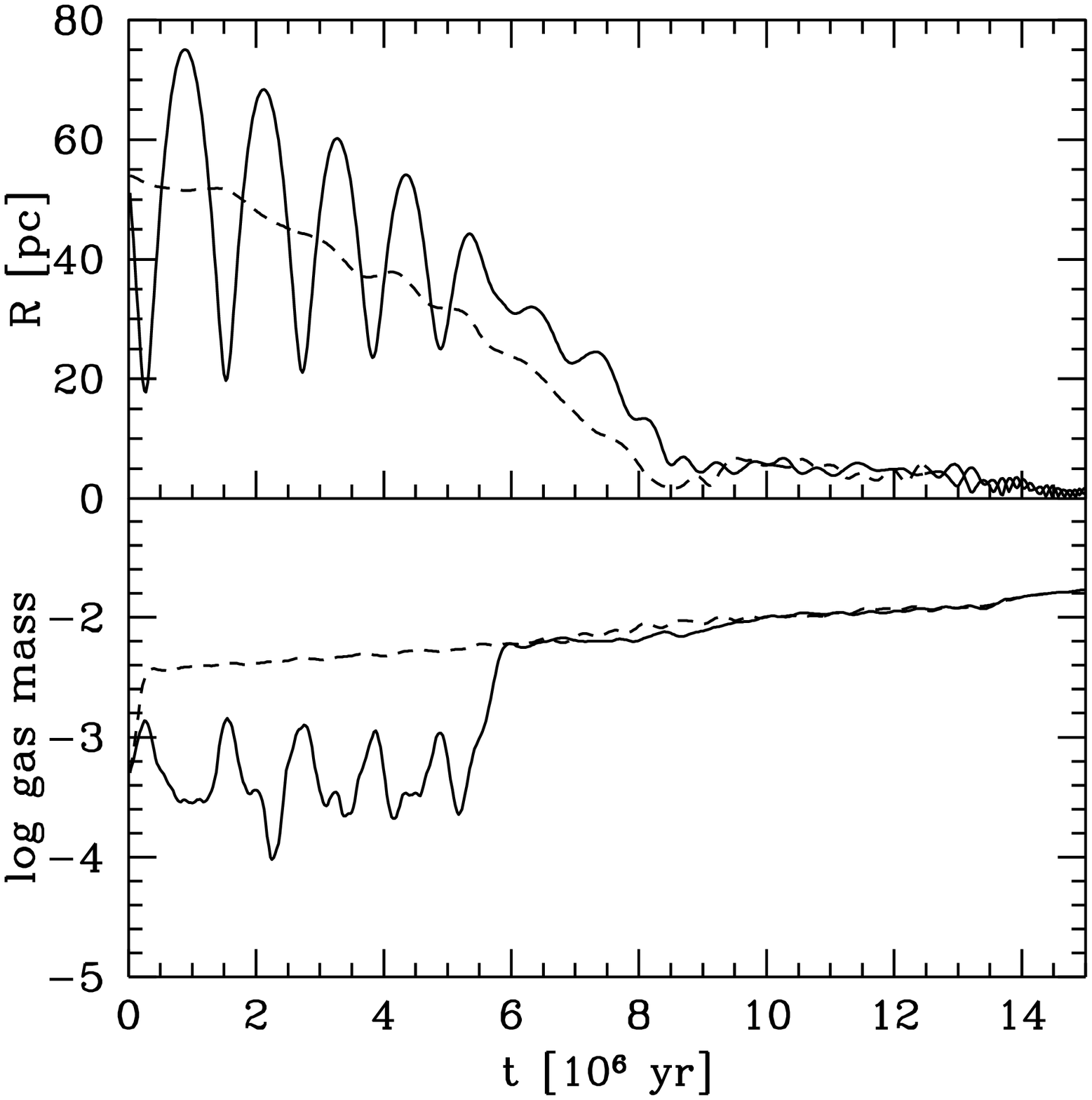,width=3.2in}}
\vspace{-0.0cm}
\caption{\footnotesize 
Upper panel: {\it Solid} ({\it dashed}) line shows the
distance $R$ (pc) of the eccentric (circular) BH from the centre of mass of the 
system as a function of time $t$.
Lower panel: {\it Solid} ({\it dashed}) line shows the mass of the 
over-density in internal units (as defined in the text)  
corresponding to the eccentric (circular) BH as a function 
of time.}
\label{ellieq}
\vspace{0.5cm}
\end{figure}

We consider here the case of two equal mass BHs in the disc plane, the
first moving on a initially circular orbit, and the second on an
initially eccentric orbit ($e=0.97$) with same binding energy (run B).
Figure~\ref{ellieq}, upper panel, shows the BH distances from the centre of
 mass,
as a function of time.  We find that the sinking time of the eccentric
BH is comparable to that of the companion, but what is remarkable is
the strong effect of circularization seen in the orbit of the eccentric
BH.  Dynamical friction in a rotationally supported gaseous medium
makes eccentric sinking orbits circular. 
This is opposite to what is
found in isotropic, purely collisionless spherical backgrounds (Colpi, Mayer \&
Governato 1999, van den Bosch \etal 1999), or in a spherical pressure--supported 
gaseous background (Sanchez--Salcedo \& Brandenburg 2000).

\begin{figure*}

\vspace{0.5cm}

\centerline{\psfig{file=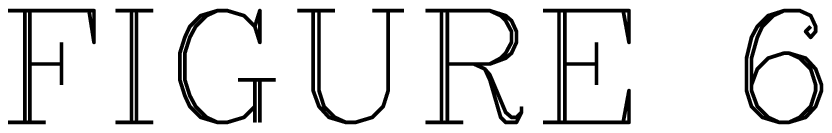,width=7.0in}}

\vspace{-0.0cm}
\caption{\footnotesize Time sequence of the sinking of the BHs, from
run B. The panels show a face--on projection of the disc and BH
positions at four different times.  The color coding indicates the z--averaged
gas density (in linear scale), and the white lines trace the BH
counterclockwise prograde orbits. In the upper left panel the
over--density created by both BHs are behind their current trail, while
in the right upper panel, the BH moving on the eccentric orbit finds
its own wake in front of its path. The wake is dragged by the faster
rotation of the disc. In the two lower panels we observe a bending of
the wake that tends to re--align, with time, behind the direction of
motion of the BH.}
\label{ellieqden}
\vspace{0.5cm}
\end{figure*}

Figure~\ref{ellieqden} shows the density map of the gaseous disc
viewed face--on.  During the sinking process the BHs develop prominent
wakes that are perturbing the underlying gas density. The BH moving
initially on a circular orbit, spirals inwards maintaining its
over-density behind its trail. The BH moving initially on the eccentric
orbit undergoes instead a remarkably different evolution.  Close to the
pericentre (upper left panel) the eccentric BH has a speed larger, in
modulus, than the local gas speed, and so its wake is excited behind
its trail.  The wake brakes the orbit and erodes the radial component
of the velocity. On the other hand, around the apocentre the BH moves more 
slowly and its tangential velocity (which dominates over the radial, at this
distance) is lower that the local rotational gas velocity. This causes
the interesting fact, clearly illustrated in the upper right panel,
that the wake is dragged in front of the BH, increasing its angular
momentum. When approaching again pericentre, the wake tends to realign
behind the BH, as shown in the two lower panels.  The net effect
highlighted in Figure~\ref{ellieqden} is the circularization of the BH
orbit (see Figure~\ref{ellieq}, upper panel).

We have run a case without disc, letting the BHs sink under the action
of the drag force due solely to the stellar bulge (run E).  

\begin{figure}
\vspace{0.5cm}

\centerline{\psfig{file=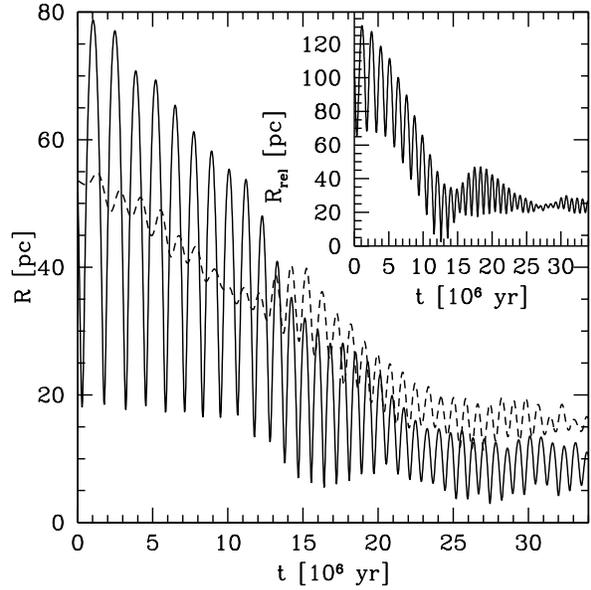,width=3.2in}}

\vspace{-0.0cm}
\caption{\footnotesize Distance $R$ from the centre of mass, as
a function of time $t$, for run E. In the simulation 
the sole collisionless bulge is considered.
{\it Solid} ({\it dashed}) line refers to the BH set initially 
onto an prograde eccentric (circular) 
orbit. In the insert we give the relative orbital
distance; axes are in the same units.
}
\label{solostelle}
\vspace{0.5cm}
\end{figure}

As illustrated in Figure~\ref{solostelle}, we find that the inspiral
takes a longer time compared to the case with gas, since the density
of stars does not rise significantly (in a Plummer model), but,
interestingly, the drag force does not lead to any circularization of
the orbit, due to the lack of rotation in the background. Furthermore,
we observe an increase in the eccentricity in the BH moving initially
on a circular orbit, in line with the findings of Colpi \etal (1999).

As far as accretion is concerned, we notice that the BH moving
along the initially eccentric orbit is unable to collect substantial
gas mass in its vicinity, given its high velocity relative to the
underlying background. Only when the orbit becomes circular the
gathering of gas can occur (see Figure~\ref{ellieq}, lower panel).
 Thus, double nuclear activity does not
always set in, but instead it depends on the properties of the BH
orbits.

\begin{figure}
\vspace{0.5cm}

\centerline{\psfig{file=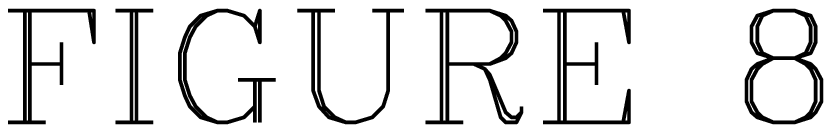,width=3.2in}}

\vspace{-0.0cm}
\caption{\footnotesize Density profile in $z=z_{\rm BHs}$ plane at time 15 Myrs. 
Lines describe isodensity regions of $\rho=0.25, 0.125, 0.09375$ and $0.0625$ (in internal 
units). Black dots correspond to the two BH positions.
}
\label{isodens}
\vspace{0.5cm}
\end{figure}

When circularization is completed, the sinking process ends as in run
A.  Figure~\ref{isodens} shows the density map, in the plane of the 
BHs $z_{\rm BH}$, at the time the two BHs form a binary system, i.e., 
when the action of dynamical friction becomes inefficient; 
an over-density of ellipsoidal shape surrounds the
binary, resulting from the superposition of the gravitational
potentials of the two BHs. 

\section{Dynamics of unequal mass black holes}

Since merging galaxies may host BHs with different masses, in this
Section, we explore the dynamics of two BHs with mass ratio 5:1 (see
Table 1 for the details of the runs). The heavier BH has a mass
$M_{\rm BH_1}=0.05 M_{\rm Disc.}$

\subsection{Circular orbits}

\begin{figure}
\vspace{0.5cm}

\centerline{\psfig{file=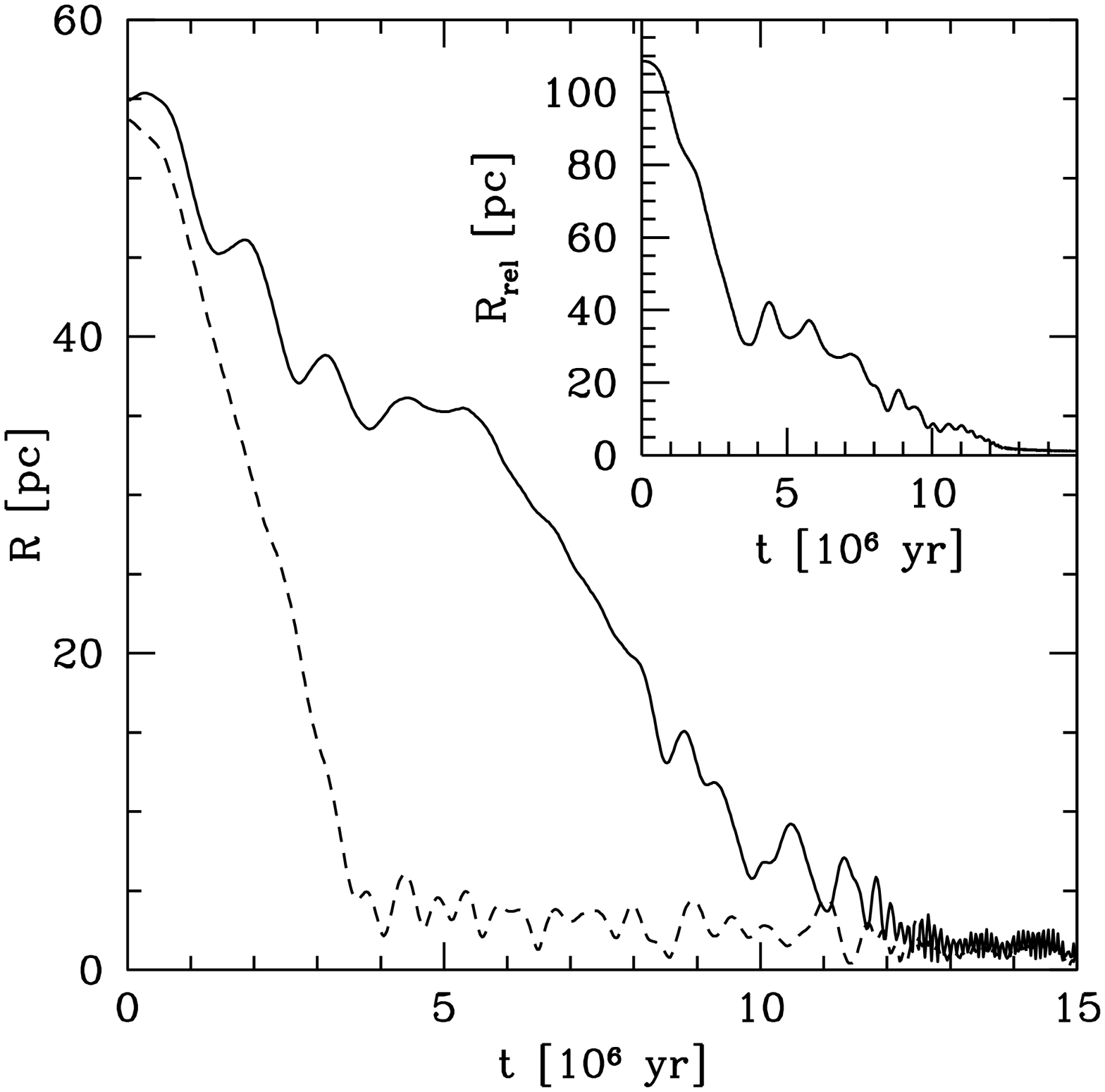,width=3.2in}}
\vspace{-0.0cm}

\caption{\footnotesize 
Same as Figure~\ref{circeq} for run C.  
{\it Solid} ({\it dashed}) line refers to the lighter
(heavier) BH.
}
\label{circuneq}
\vspace{0.5cm}
\end{figure}

\begin{figure}
\vspace{0.5cm}

\centerline{\psfig{file=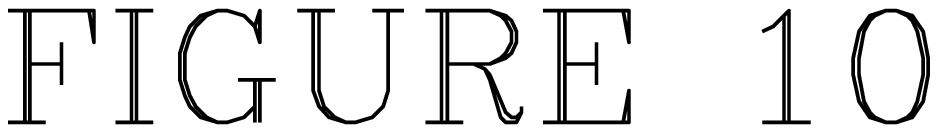,width=3.2in}}

\vspace{-0.0cm}
\caption{\footnotesize Same as Figure 6 for run C. In the upper panel 
($t=$ 1 Myr) the heavier BH perturbs the gaseous disc and shifts 
the barycentre of the system away from the lighter one. In the lower panel 
($t=$ 3.5 Myr), the heavier BH is located between the other BH and the
barycentre, speeding up the orbital decay of the lighter one.}
\label{circuneqden}
\vspace{0.5cm}
\end{figure}

We first explored the case in which the two BHs move initially on
circular prograde orbits, at equal distances from the centre of mass
(run C).  The large, more massive BH sinks rapidly toward the centre
of the gaseous disc, over a timescale of $\sim 4$ Myrs, as illustrated
in Figure~\ref{circuneq}, while the lighter completes its orbital
decay on a timescale longer by a factor $\simeq 2-3$.  This implies
that the dynamical friction time does not scale exactly as the inverse
of the mass, and we interpret this result as an effect related to the
perturbation in the overall disc gravitational potential caused by the
larger BH.  During the early stages, the two BHs are
symmetrically displaced relative to the centre of mass, and start to
develop their own wakes of different intensity.  In the upper panel of
Figure~\ref{circuneqden} we catch the instant at which the heavier BH,
having perturbed the gaseous mass, causes a shift of the barycentre,
i.e., a displacement in the direction opposite to the lighter BH.  At
this time, the orbital decay of the less massive BH halts
temporarily (as shown also in Figure~\ref{circuneq} between 3 and 6
Myrs).  When the orbital decay of the heavy BH is sufficiently
advanced that it has almost reached the centre, its location is in
between the position of the lighter BH and the barycentre
which is now  shifted toward the light hole deepening the potential
well. This causes an acceleration toward the
disc centre speeding the orbital decay (see lower panel of
Figure~\ref{circuneqden} and Figure~\ref{circuneq} at times $\gtrsim
6$ Myrs).  This case illustrates that time variations in the underlying
gravitational potential, that can be computed self--consistently in
a real simulation, can modify granted dependences of the dynamical
friction timescales. 
We find also that dynamical friction is effective until 
we hit the force resolution limit, and that no ellipsoidal density 
distribution is found (see \S 4.2 for a discussion).

Our initial condition is clearly arbitrary, and it is likely that in
real mergers the heavier BH is already in place at the centre of the
circum-nuclear disc by the time the second BH enters the disc. For this
reason the sinking process may be different for the same BH masses
involved, depending on the details of the process of paring.  In the
next simulation where the lighter BH is set onto an eccentric orbit, we
allow the more massive hole to reach the centre before the sinking
process of the light one takes place.

\subsection{Eccentric orbits}

\begin{figure}
\vspace{0.5cm}

\centerline{\psfig{file=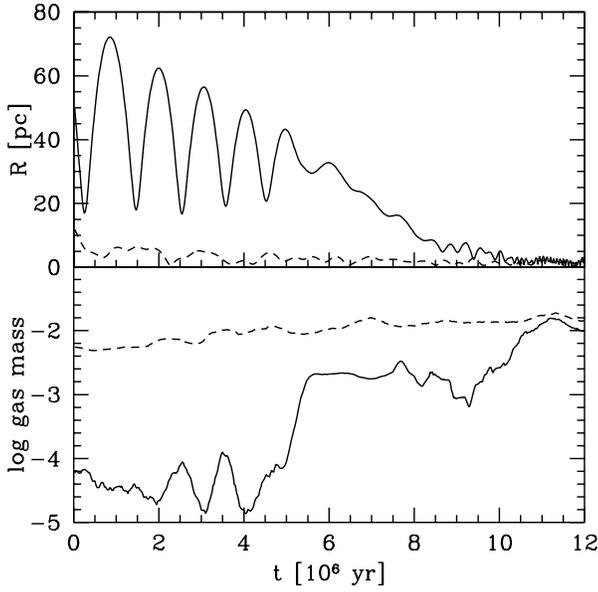,width=3.2in}} 

\vspace{-0.0cm}
\caption{\footnotesize Upper panel: {\it Solid} ({\it dashed}) line shows the
distance $R$ (pc) of the lighter (heavier) BH from the centre of mass of the 
system as a function of time $t$.
Lower panel: {\it Solid} ({\it dashed}) line shows the mass  (in internal 
units)of the  over-density  corresponding to the lighter (heavier) BH as a
 function of time.}
\label{pippo}
\vspace{0.5cm}
\end{figure}

\begin{figure}
\vspace{0.5cm}

\centerline{\psfig{file=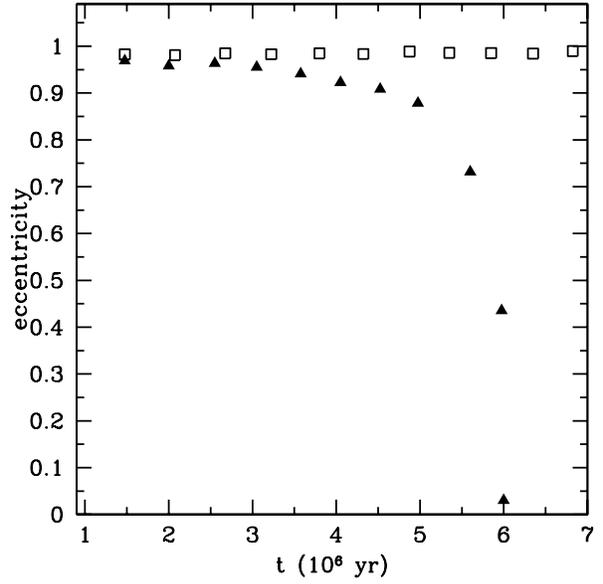,width=3.2in}}

\vspace{-0.0cm}
\caption{\footnotesize Eccentricity of the less massive BH as a function of time, in the case of prograde 
(Run D, {\it triangles}), and retrograde (Run F, {\it squares}) orbit. The eccentricity is computed 
over any half orbit.   
}
\label{gilberto}
\vspace{0.5cm}
\end{figure}

\begin{figure}
\vspace{0.5cm}

\centerline{\psfig{file=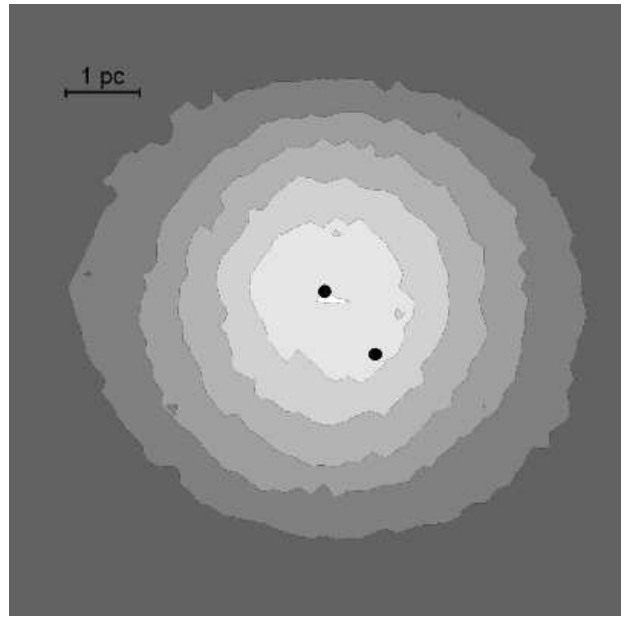,width=3.2in}}
\vspace{-0.0cm}
\caption{\footnotesize Same as Figure 8 for run D at time 13 Myr. 
The lines describe regions of $\rho = 2, 1, 0.5, 0.25, 0.125, 0.0625$ (in internal units).
The isodensity contours are almost circular.}
\label{isodensunoacinque}
\vspace{0.5cm}
\end{figure}

In run D, the light BH is initially set on a prograde orbit with
$e=0.95$.  Its sinking as a function of time proceeds, similarly to
run B, with the circularization of the orbit.  Since the larger BH is
already in place at the centre of the gaseous disc we do not see any
effect on the acceleration of the orbit besides dynamical friction.
As shown in the upper panel of Figure~\ref{pippo} the orbit of the
light hole becomes circular at a time $t\simeq 5$ Myrs. 
The evolution of the eccentricity is shown in Figure~\ref{gilberto}.
In the early inspiral, the gas mass associated to the over-density in the
neighborhood of the small BH is only $M_{\rm Gas}\sim 3\times
10^{-3}M_{\rm {BH_l}}$, and only when the orbit becomes circular, we
observe a rapid increase to a value $\sim 0.25 M_{\rm BH_l},$
potentially triggering an episode of accretion.  When the light hole
binds to the larger one ($t\simeq 11$ Myrs), it finds itself embedded
inside the over-density created by the large hole, as shown in the
lower panel of Figure~\ref{pippo}.  Contrary to the case of equal mass
BHs, at very late times, the overdensity distribution around the
binary BHs is not any longer ellipsoidal in shape and the
gravitational field is weakly dipolar.  Figure~\ref{isodensunoacinque}
shows how strong is the degree of sphericity of the gas surrounding
the two BHs.  Does orbital decay proceed further?  The light BH seems
to decay but at a much lower pace.  The lack of a visible wake and
the absence of an ellipsoidal deformation (torquing the binary
components) suggest that the gas has become sufficiently stiff and
the potential well sufficiently deep, due to the presence of the more
massive hole, that orbital decay is halted or delayed.

\begin{figure}
\vspace{0.5cm}

\centerline{\psfig{file=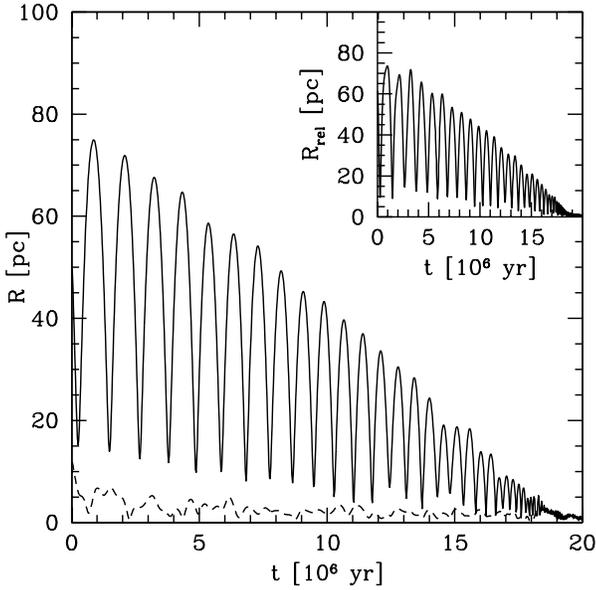,width=3.2in}}

\vspace{-0.0cm}
\caption{\footnotesize Same as Figure 3 for run F. {\it Solid} ({\it dashed}) 
line refers to the lighter
(heavier) BH.}
\label{sisisi}
\vspace{0.5cm}
\end{figure}

In run F, we explore the dynamics of an unequal mass BH
set on a retrograde eccentric orbit. This is opposite to run D and
is considered in order to bracket uncertainties in the way
BHs bind. We find that the orbit remains eccentric (see Figure~\ref{gilberto}),
preventing the gas to accumulate substantially around the BHs during
 the whole inspiral process.
Given the rotational pattern
of the gas, the over-density created by the BH stays always
behind its trail, so that the eccentric BH does not circularize, as shown 
in figure~\ref{sisisi}. Note that the sinking time is about twice larger if compared 
to the prograde cases. 

\section{Discussion}

We explored the dynamics of a MBH pair orbiting inside
a gaseous disc embedded in a spherical stellar distribution. We
followed the slow inspiral of the pair, driven by dynamical friction,
until the MBHs bind to form a close binary.  The calculation is
idealized in many ways, since we assumed a particular density distribution 
for the gas particles (Mestel disc), and neglected gas cooling and star
formation.

Despite these limitations, we highlighted basic features
of the dynamics of {\it LISA} double BHs in gas--rich discs.  When a
BH initially is moving on a highly eccentric co--rotating orbit, its
eccentricity decreases significantly, contrary to what occurs when the
background is spherical and collisionless (as shown in Figure~\ref{solostelle};
 Colpi et al. 1999).  Disc rotation is the key element of the circularization:
near apocentre, where the angular velocity of the MBH is smaller than
that of the disc, the density pattern, created by the MBH along its
motion, is dragged in front of the BH itself enhancing its angular
momentum.  As a consequence the MBHs tend to form a close binary with
a low eccentricity. However, the numerical noise at the end of the
simulation does not allow us to calculate the precise value of any
residual eccentricity.  In the case of counter--rotating orbits, the
eccentricity does not decrease since the density wake remains always
behind the BH motion, and the MBHs may end forming a binary with still
significant eccentricity (Figure~\ref{gilberto}). 
The probability of pairing along counter or
co--rotating orbits is not known yet, and should be further
investigated creating a statistically significant sample of
simulations of gas--rich merging galaxies, similar to those carried out by
Kazantzidis \etal (2005).
 
To summarize, if co--rotating orbits are more likely, we can note that
the braking of MBHs in a gaseous background deliver a MBHB on a
nearly circular orbit. It is worth noticing that, during later phases, further eccentricity
evolution may still occur, driven by
close encounters with single stars 
(Mikkola \& Valtonen 1992, Quinlan 1996, Milosavljevic \& Merritt
2001, Aarseth 2003, Berczik, Merritt \& Spurzem 2005), and/or 
gas--dynamical processes (Armitage \& Natarajan 2005),
before GW emission acts to circularize the orbits.
In Armitage \& Natarajan (2005)
the eccetricity is shown to grow substantially at very small separations, 
when gravitational torques from the MBHB act to clear a gap in the circum--binary disc.
This fact may have important consequences for a possible final, GW driven, coalescence of 
the two BHs.  

We find in addition, that when dynamical friction has subsided, in
the case of equal mass BHs, the binary that forms is surrounded by
gaseous particles belonging to the head of the wakes that merge in a
coherent pattern, i.e., an ellipsoidal mass distribution. 
In the case of unequal masses, we observe on the contrary
that the trailing over-density created by the light BH brings it to a
closer distance from the heavier central BH.  Given the large
unbalance between the two masses, the gas distribution around the
binary remains remarkably spherical and there is no gravitational
torque in action to cause further inspiral of the light BH, within our
resolution limit.

The feeding of BHs during their inspiral is also a key related issue.
Despite the limitations of the thermodynamics employed, we have shown that 
a MBH moving on an eccentric orbit is unable
to gather a relevant amount of gas in its vicinity, having a much higher
velocity relative to the background. Only when the MBH orbit
becomes nearly circular, gas is collected close by. This may conduct
to the formation of an accretion disc fueling the MBH, hence triggering 
nuclear activity, observable on scales of a few pc, during the pairing.
Then, AGN activity could be linked to the dynamics of the
pairing process of the MBHs inside circum--nuclear discs.  
The possible presence of accretion discs near the horizons of coalescing MBHs is of particular 
importance, as it may leave an electromagnetic signal correlated the GW emission targeted by 
{\it LISA} (Armitage \& Natarajan 2002, Milosavljev\'ic \& Phinney 2005, Kocsis \etal 2005).
We plan to improve upon our model, e.g., including gas cooling and heating, star formation, BH 
treated as sink particles, in order to explore the accretion issue in much greater detail. 

Whether circularization stalls the {\it LISA} BHs to separations
larger than critical for the intervention of GWs is not clear
yet. The exploration of later stages requires a much higher force
resolution and a modeling of the BHs as ``absorbing'' particles,
having an horizon, i.e., a trapping surface.  ELCM05 show, for two
$M_{\rm BH}=5 \times 10^7 \, \msun$, that the dynamical action of the
torque continues at least down to a separation $\simeq 0.1$ pc, where
the time to coalescence (because of gravitational wave emission) is
about $10$ Gyrs. However it is not clear if the process is still
relevant for different conditions, not discussed by ELCM05, i.e.,
different disc--to--binary mass ratio, and/or different BH--BH mass
ratio.  We can just note that binary BHs in gaseous rotating
background do not bind on those highly eccentric plunging orbits that
would bring them straight to coalescence. {\it LISA} BHs form circular
binaries and may need additional mechanisms for continuing their
hardening process. Three--body encounters with low angular momentum
bulge stars is a possibility (Mikkola \& Valtonen 1992, Quinlan 1996),
though it seems that BH binaries must be on highly eccentric orbits to
drive the separation to sub-pc scales, where gravitational wave
emission takes over (Sesana \etal 2005, in preparation).  One aspect worth
considering is that, during the hardening phase, one or both BHs could
accrete gas, hence increase their mass, modifying the dynamics of the
binary.  Lets us now suppose that the mass of the member $M_1$ ($M_2$) of
a circular binary increases by a factor $\beta_1$ ($\beta_2$). Then, if 
the binary  
angular momentum is conserved, we have a reduction of the binary
separation by a factor
\be
\frac{a_{\rm new}}{a_{\rm old}}=\frac{\beta_1 M_1 + \beta_2 M_2}{\beta_1^2 \beta_2^2(M_1+M_2)}.
\label{accretion}    
\ee
As an example, let us consider a very unequal mass binary ($M_1 \gg M_2$):
the heavier BH can indeed accrete while the secondary is slowly
spiraling inwards, as in run D.  
From equation \ref{accretion}, an increase of $M_1$ by a factor of, say, 2, 
reduces the binary separation by the same factor. 
This is quite promising, as, in scattering 
experiments (Sesana \etal 2005, in preparation), {\it LISA} circular binaries tend to 
stall at separations which are factors $\sim 2-8$ (depending on mass and mass ratio) 
too large to drive the binary to coalesce within 1 Gyr because of gravitational wave emission. 

\section*{acknowledgments}
The authors thank Andres Escala, Lucio Mayer, Ruben Salvaterra, Javier Sanchez, 
Boris Sbarufatti and Alberto Sesana for fruitful 
comments and suggestions, and Franz Livio and Luca Paredi for technical support.

{}

\end{document}